# Parallel Genetic Algorithm to Solve Traveling Salesman Problem on MapReduce Framework using Hadoop Cluster

Harun Raşit Er
Faculty of Computer and Informatics
Istanbul Technical University, Turkey
hrer@itu.edu.tr

Prof. Dr. Nadia Erdoğan
Faculty of Computer and Informatics
Istanbul Technical University, Turkey
nerdogan@itu.edu.tr

*Abstract*- Traveling Salesman Problem (TSP) is one of the most common studied problems in combinatorial optimization. Given the list of cities and distances between them, the problem is to find the shortest tour possible which visits all the cities in list exactly once and ends in the city where it starts. Despite the Traveling Salesman Problem is NP-Hard, a lot of methods and solutions are proposed to the problem. One of them is Genetic Algorithm (GA). GA is a simple but an efficient heuristic method that can be used to solve Traveling Salesman Problem. In this paper, we will show a parallel genetic algorithm implementation on MapReduce framework in order to solve Traveling Salesman Problem. MapReduce is a framework used to support distributed computation on clusters of computers. We used free licensed Hadoop implementation as MapReduce framework.

I. INTRODUCTION

The Traveling Salesman Problem (TSP) is a well-known and important combinatorial optimization problem. The goal is to find the shortest tour that visits each city in a given list exactly once and then returns to the starting city. In contrast to its simple definition, solving the TSP [1] is difficult since it is an NP-complete problem. It has many application areas such as planning, logistic and network routing, chip design and manufacturing.

There are numerous approaches to solve TSP. In this paper, we examine Genetic Algorithm (GA) solution. GA is a heuristic solution which yields near optimal solutions within a reasonable time. Although GA is time efficient and a good approximation for TSP, sometimes it can stuck to local optima or it takes considerable time when the number of cities increases.

The inherent parallel nature of evolutionary algorithms makes them optimal candidates for parallelization [2]. There are many studies on parallelization of Genetic Algorithms on a computer cluster. MPI-based parallelization is one of the most studied methods. MapReduce [3, 10] parallelization is another way of parallelization which is studied in this paper.

MapReduce is a framework that is comprised of map and reduce functions. It enables users to develop large-scale and fault tolerant distributed applications. Applications developed on MapReduce framework are naturally self-fault tolerant. Hadoop [8] is a software implementation of MapReduce framework. Hadoop runs on a commodity hardware cluster which is much cheaper than a specialized workstation. Its base approach is to transfer the program code to the data node instead of transferring data across the network. As a result, it overcomes the data transferring bottleneck of the distributed applications.

In the next section, we give a detailed description of the TSP and its application areas. The third section covers the Genetic Algorithm and how it can be utilized as a solution for TSP. Later we describe parallelization methods of GA and how GA can be expressed as a MapReduce job. At the conclusion of paper, we report our experimental results and comparisons to related works.

II. TRAVELING SALESMAN PROBLEM

Traveling Salesman Problem is one of the most studied combinatorial problems because it is simple to comprehend but hard to solve [5]. The problem is to find the shortest tour of a given number of cities which visits each city exactly once and returns to the starting city [4].

In a complete weighted undirected graph G (V, E) where cities are represented by vertices and distances are presented by weighted edges, TSP is to find the minimized Hamilton cycle that starts from a specified vertex, visits all the other vertices exactly once, and ends at the same specified vertex.

At the first glance, TSP seems to be limited for a few application areas; however it can be used in a lot of problem solutions. Some of the application areas are; printed circuit manufacturing, industrial robotics, time and job scheduling of the machines, logistic or holiday routing, specifying package transfer route in computer networks, and airport flight scheduling.

As a solution of TSP, there are two main approaches. The first tries to find an optimal solution that guarantees the quality of solution; however it is very slow and mostly infeasible for larger problem sizes. The second one tries to find a solution within a reasonable time without any guarantee for an optimal solution [6]. Its goal is to get better performance with a lack of optimality. Our approach in this study is to find a near optimal solution in an acceptable time.

The exact solution would be to try all permutations and choosing the cheapest one using brute-force search. Even though this method guarantees the best solution for a small number of cities, it becomes impractical even for 20 cities.

### III. GENETIC ALGORITHM FOR TSP

Genetic Algorithms are well suited for combinatorial optimization problems. Because of their ease of use, and efficient results, they are very popular in solving NP-Hard problems. The algorithm is inspired from natural evolution where always the best individuals have a chance of survival.

The algorithm has five main stages. Firstly, an initial population is generated. Next, fitness of all individuals in the population is evaluated. According to the fitness values of individuals, good ones are selected for the reproduction of the new population. After selection operation, crossover between selected individuals is carried out to produce the next generation. The last step is to mutate individuals with a given probability for diversity of search directions. If the resulting generation has a solution which is optimal, then the algorithm is terminated. Otherwise, fitness evaluation, selection, crossover and mutation operators are applied on the next generation iteratively.

Creating an initial population is the first step of a genetic algorithm. For TSP, it means creating a list of tours which visit every city exactly once and return to starting city.

While creating a tour – it is a solution for the problem – another consideration is encoding the solution. In our work, permutation encoding is used. Every city has an integer identifier that ranges from 0 to $N$ where $N$ is the problem size. After one of the cities is picked up as starting city, another city among the unvisited cities is selected until all cities are visited.

After the initial population is created, the next step is fitness evaluation of individuals. First, the path length of each tour is calculated. After the calculation of path length of all individuals, the fitness value of each individual is assigned by dividing its path length by total path length.

Selection operator is applied to individuals in the population. Rank selection with elitism is used as a selection operator. Rank selection ranks the population and every individual receives a rank value. The worst one has a rank of 1; the second worst a rank of 2 and so on. The best individual has a rank of $N$ which is the number of individuals in the population. Next, each individual is assigned a probability to be chosen for crossover operation in order to generate next population.

One of the drawbacks of the genetic algorithm is that it can get stuck in local optimum. In order to avoid this situation, we propose preventing consanguineous marriage. Our policy in this respect is selecting two individuals that have different gene orders over a particular percent for the crossover operation. Setting a percentage value too high prevents the algorithm from converging. Otherwise, if the percentage value is minimized too much, it has no effect on avoiding from local optima. After some experimentation, we have chosen the percentage value as %20. Therefore, if two parents have similarity over %80, they cannot be chosen for the crossover.

The next stage is to make crossover between selected two parents. Several crossover methods have been proposed for TSP. Some of the most common methods are Partially Mapped Crossover, Order Crossover and Edge Crossover. In this study, we used the Greedy Crossover method.

The Greedy Crossover procedure is simple. For a pair of parents, one of the cities is picked up as a starting city; the shortest edge that is presented in the parents, leading from the current city and not introducing a cycle is chosen. If the shortest edge leads to a cycle, the other edge from the other parent is chosen. On the other hand, if both edges from parents lead to a cycle then we randomly choose a city that does not lead to a cycle. In this manner, selection of subsequent cities continues until the tour is completed.

When the crossover stage is completed, a mutation operation is applied to individuals in the next generation with a certain probability which is specified as 2.1% in this study. In the mutation process, randomly chosen two genes are replaced with each other.

When the next population is produced, it is determined whether convergence is obtained or not. If the solution satisfies the termination condition, the algorithm is terminated. Otherwise fitness evaluation, selection, crossover and mutation stages are repeated until convergence is obtained.

### IV. PARALLEL GENETIC ALGORITHMS

As mentioned above, Genetic Algorithm is comprised of fitness evaluation, selection, crossover and mutation phases. The most time consuming stage in the algorithm is fitness evaluation. Especially if the population size is very large, fitness evaluation turns into a big problem to be solved in sequential GA. Another problem encountered in a sequential genetic algorithm is that it sometimes gets stuck in a local search space [7].

To overcome the problems described above, Parallel Genetic Algorithm (PGA) is used. Fitness evaluation problem can be solved using multi-core or super computers. If getting stuck in sub-optimal search space is the problem, PGAs provide an appropriate solution for the problem.

The calculation of fitness evaluation, the use of single or multiple population, in case of multiple populations, the way individual exchange is carried out are some of the criteria according which PGAs are classified. PGAs are classified into Master-slave, multiple populations with migration and multiple populations without migration.

In master-slave parallelization, the most time consuming part of the genetic algorithm, that is fitness evaluation, is calculated in parallel. In this class of PGAs, all stages of the genetic algorithm except for fitness evaluation are performed on a master node in a sequential manner. Master node sends the individuals to the slave nodes, which calculate the fitness of the given individuals and return the results back to master node. When the master node gets the fitness of all individuals in the population, selection, crossover and mutation operations are applied globally.

Even though Master-slave parallelization outcomes the fitness evaluation problem, the other parts of the algorithm are still processed sequentially. Also, the local search problem still remains.

Multiple populations parallelization method which is the mostly applied method, uses multiple sub-populations [7] as the name indicates. All the genetic operators are applied to sub-populations separately. Sub-populations evolve independently of each other.

In multiple populations without migration parallelization method, some individuals of the sub-populations are shared between each other. So search direction traced by each sub-population is transferred to other sub-populations.
Multiple populations with migration parallelization method is the method we have used. It uses migration operator in order to share search direction between sub-populations. Some of the individuals from one sub-population are transferred to other sub-populations at regular intervals [7].

Migration operator has some parameters that need to be optimized: The interval that migration operator is applied; the number and characteristics of individuals transferred and replaced should be determined.

Migration interval determines how frequently the selected individuals are transferred to other sub-populations. As stated above, sub-populations evolve separately and some selected individuals are migrated to other sub-populations at determined intervals.

While migrating individuals, selecting the kinds of chromosomes to be transferred and specifying the kinds of chromosomes in the other sub-population with which they are to be replaced is an important process. In this study, the best chromosomes are selected and migrated to other sub-populations. Accordingly, the worst chromosomes in other sub-populations are replaced with the newly migrated chromosomes.

V.  MAPREDUCE FRAMEWORK

MapReduce is a distributed computing framework proposed by Google for processing large data sets on a cluster. It enables users to develop and run distributed programs easily.

Network bandwidth bottleneck is the most encountered problem in distributed applications [8]. MapReduce framework overcomes the network bandwidth bottleneck through data locality that is by collocating running code and data.

Another problem in distributed applications is failure of a node or failure of connection to a node. MapReduce has self failure detection and recovery procedure. Hence, a developer concentrates totally on its application with no concern on failures.

MapReduce framework runs in slave-master model. In traditional master-slave programming models, a developer should consider the coordination of nodes. On the other hand, MapReduce framework handles coordination by itself, using RPC calls between master and slave nodes.

MapReduce framework, as the name indicates, consists of map and reduce functions which are written by the user. Map function reads key-value pairs from file system, groups them according to their keys and creates intermediate key-value pairs [3]. Reduce function receives a key and a list of values that are associated with the key. It performs a set of operations and the resulting key-value pairs are written back to file system by reducer.

Hadoop [11], free licensed implementation of MapReduce architecture, is used in this study as a MapReduce framework. It provides two main concepts; computational architecture for MapReduce jobs and Hadoop Distributed File System (HDFS) [12] for input and output data of MapReduce jobs. Hence, Map functions read their inputs from HDFS and create intermediate key-value pairs. Reduce functions receive intermediate key-value pairs and after performing certain operations they write results back to HDFS.

Hadoop cluster [11] has five main components, namely NameNode, DataNode, Secondary NameNode, JobTracker, and TaskTracker. NameNode is responsible for managing data on HDFS. DataNode, as the name suggests, stores data and interacts with NameNode. Secondary NameNode runs as a back-up for NameNode. TaskTracker communicates with client and runs the client's MapReduce jobs via TaskTracker and coordinates TaskTrackers to complete job consistently. TaskTracker is responsible for the execution of map and reduce tasks which constitute the MapReduce job. NameNode, Secondary NameNode, JobTracker are the master part of MapReduce framework and they all run on a master machine. DataNode and TaskTracker are slave components which run on slave machines. The Hadoop cluster has one NameNode, one Secondary NameNode and one JobTracker, while the cluster can have any number of DataNodes and TaskTrackers.

## VI. METHOD – MAP-REDUCING GA

Static population with migration parallelization method is used to parallelize the genetic algorithm. Sub-populations evolve on their own. And they exchange good individuals between each other at regular intervals as shown in Fig. 1.

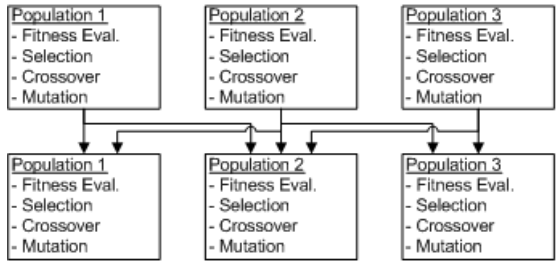

**Figure 1: Static populations with migration**

Iterative MapReduce is used to implement Parallel GA. Each evolution phase is implemented as a MapReduce job. Each job has N number of map and reduce tasks where *N* is the number of sub-populations. At the end of a job, individuals are written back to HDFS file system as shown in Fig. 2. Before starting the next evolution phase, it is checked out whether convergence has occurred or not. If it does, client terminates the program and the result is presented as an optimal solution. Otherwise, client starts the next MapReduce job and the sub-populations develop until the next exchange phase.

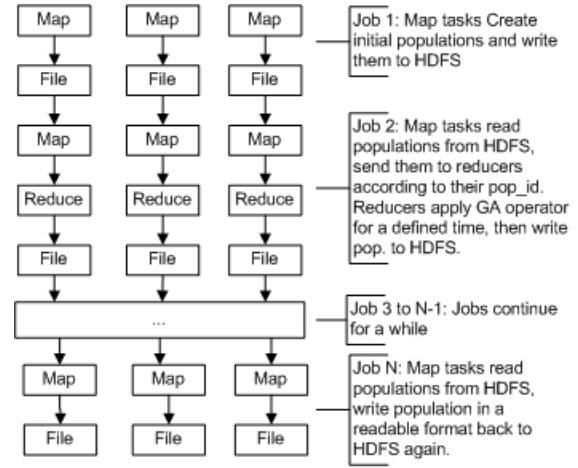

**Figure 2: MapReduce implementation of static populations with migration**

MapReduce framework uses string class for value and integer for key, which are not convenient for GA. Therefore, we have implemented our own chromosome input and output format. Our *Chromosome* class implements Hadoop's *Writable* interface, and overrides *readFields* and *write* functions of the interface. As a result, individuals are directly written to and read from HDFS without any interpretation.

The main components we have implemented for our MapReduce application are Driver class, Mapper function, Reducer function and Partitioner class which directs individuals that share the same population identifier to the same reducer/population.

*Driver Class*
BEGIN
   Run Job1: Creates initial population in parallel
   FOREACH max generation number
     Run Job *i*: evolve population
   END FOR
   Run job N: write resulting populations HDFS in
   a readable format to HDFS
END

Driver class is the program entry point. Client sends jobs to JobTracker, and decides to terminate a program via this class. It is also used to set the initial Hadoop environment parameters such as the number of map and reduce tasks, input/output format and directory, etc.

Map function is used to read individuals from HDFS file system with their population identifier and group individuals according to their population identifier.

Hadoop has its default partitioner as hash partitioner which produces the partition number using chromosome's gene order in our case. So it sends the individuals which have the same gene order to same reducer (sub-population). However, this approach is not convenient for our problem. Therefore, we replace it with our partitioner which shuffles individuals according to their population identifiers. As a result, all individuals that share the same population identifier are sent to same reducer.

Reducer receives the individuals that belong to same population. Population evolves for a specific iteration number. Firstly, rank selection is applied to select suitable parents for crossover. Using greedy crossover, a new population is obtained. Next, mutation is applied for recently generated individuals with a certain probability. After evolution for determined iteration, the new population is written back to HDFS file system. The best individuals in each population are written with different population identifiers. So, for the next iteration, they are sent to other populations.

Each sub-population is assigned to a different reduce task and reduce functions evolve sub-populations until migration process starts.

*Reducer Class*
BEGIN
  Receive individuals that all have the same id
  FOREACH evolution number
    FOREACH population size
      Apply rank selection
      Apply Greedy crossover
      Add new individuals to new population
    END FOR
    Apply mutation to new population
  END FOR
  Write population to HDFS
  FOREACH number of other sub-populations
    Change individual pop_id and write it to HDFS
  END FOR
END

After iteration has completed, all reducers write the best individual in their sub-populations to the file system. Before starting the next evolution step, client program reads the best individuals and decides if the convergence criteria have been satisfied.

## VII. RESULTS AND EVALUATION

To assess the performance of our parallel implementation, we compare it with a sequential implementation of the genetic algorithm. The sequential genetic algorithm (SGA) we have developed is also compared with other implementations [9].

Sequential GA runs on Intel Core Duo 2.4 GHz machine including 3GB of RAM. Parallel GA runs on a Hadoop cluster of 6 machines that are described on Table 1.

| Name | CPU (GHz) | Core | CPU Type | RAM (GB) |
|---|---|---|---|---|
| Master | 3.0 | 2 | P4 | 2.0 |
| Slave1 | 3.2 | 2 | P4 | 2.0 |
| Slave2 | 3.0 | 2 | PD | 2.0 |
| Slave3 | 2.13 | 2 | Duo | 3.0 |
| Slave4 | 2.33 | 4 | Quad | 3.5 |
| Slave5 | 2.8 | 4 | I5 | 2.5 |

Table 1: Machines on Hadoop Cluster

The problem instances we have used in this study are taken from TSPLIB library [13]. The problems are asymmetric, that is, the distance from node *i* to node *j* and the distance from node *j* to node i could be different. The results are obtained by execution of the algorithms for at least 10 times. Population size of sequential GA and each sub-population size of MapReduce GA are set to 100. The other GA parameters; crossover probability and mutation probability are set to %99 and %2.1 respectively.

SGA is compared with Edge Recombination crossover (ERX), Generalized N-Point crossover (GNX) and Sequential Constructive crossover (SCX) that are implemented in [9]. Time and optimal solution accuracy comparison of the algorithms are shown in Fig. 3 and Fig. 4 respectively. For this experiment, maximum generation number is set to 10.000.

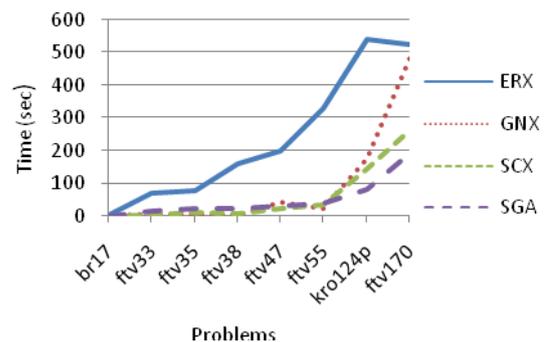

Figure 3: Solution Time for ERX, GNX, SCX and SGA in seconds for TSPLIB instances

The time requirements of four algorithms are shown in Fig. 3. ERX is the most time consuming algorithm. GNX algorithm takes the same with SGA for small problem sizes. For larger problem sizes, it takes more time than SGA. SCX and SGA take almost the same time for the solutions.

Results in Fig. 4 show that the SGA finds the optimal solution faster than the others always. Also it finds the better solution than the other algorithms.

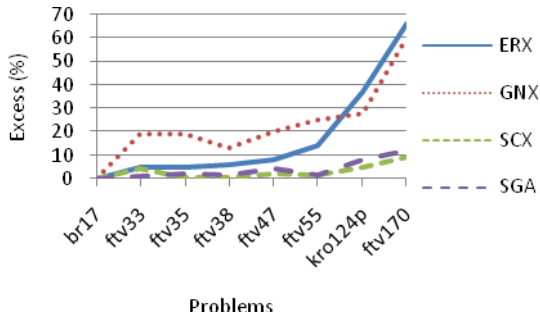

Figure 4: Average solution accuracy of algorithms as percentage for TSPLIB instances

Next, we compare SGA with the parallelized MapReduce GA. We set maximum generation number as 50.000 in this experiment.

Since each sub-population in MapReduce GA searches solution space in a different direction, MapReduce GA always finds better solutions than SGA Fig. 5.

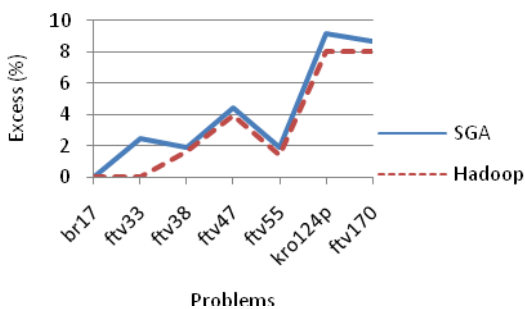

Figure 5: Percentage of average solution accuracy for TSPLIB instances

Fig. 6 shows the time required for both sequential and MapReduce GA. Sequential algorithm obtains solution faster than MapReduce GA for small sized problems. This is true because JVM creation time for map and reduce tasks dominates the solution time. However, when the problem size increases, sequential GA solution time increases dramatically. MapReduce GA has almost the same run time for all problem sizes and it hammers the sequential GA for large problem sizes.

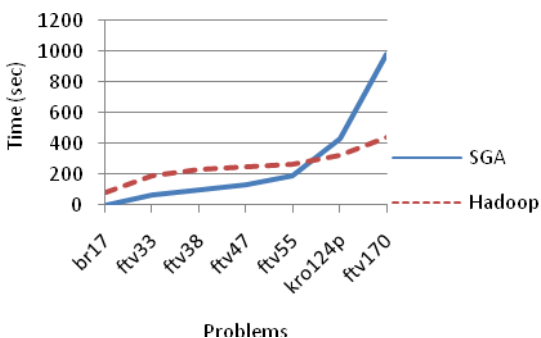

Figure 6: Solution Times of SGA and HADOOP in seconds for TSPLIB instances

## VIII. CONCLUSION

We used genetic algorithm in order to solve Traveling Salesman Problem. And we parallelized the algorithm on Hadoop Cluster. Static population with migration method is used as the parallelization method.

We compare sequential GA with other studies. Our sequential GA always gives better solutions than the others in terms of quality and time.

MapReduce parallel genetic algorithm comparison with sequential genetic algorithm shows that MapReduce GA finds better solutions and takes shorter time than SGA when the problem size increases.

The maximum problem size used in this experiment is 171. Even though, this study shows that the Hadoop parallelization gives better results than sequential algorithm, we consider examining still larger problem sizes to compare results with other parallel implementations. Using a larger Hadoop cluster is in our future work plan.

We used 10 sub-populations that is, 10 map/reduce tasks run parallel for each Job. Increasing or decreasing the number of tasks may also reveal interesting results.

## IX. REFERENCES

[1] E. L. Lawler, J. K. Lenstra, A. H. G. RinnooyKan, and D. B. Shmoys (1985), The Traveling Salesman Problem, John Wiley & Sons,Chichester

[2] E. Cant´u-Paz., Efficient and Accurate Parallel Genetic Algorithms. Springer, 2000

[3] Abhishek Verma, Xavier Llor`a, David E. Goldberg and Roy H. Campbell, Scaling Simple and Compact Genetic Algorithms using MapReduce in International Conference on Intelligent Systems Design and Applications, 2009

[4] Fan Yang, Solving Traveling Salesman Problem Using Parallel Genetic Algorithm and Simulated Annealing, 2010

[5] Adewole Philip, Akinwale Adio Taofiki, Otunbanowo Kehinde, A Genetic Algorithm for Solving Travelling Salesman Problem, in International Journal of Advanced Computer Science and Applications, 2011

[6] Siddhartha Jain, Matthew Mallozzi, Parallel Heuristics for TSP on MapReduce, Brown University, 2010

[7] Mariusz Nowostawski, Riccardo Poli, Parallel Genetic Algorithm Taxonomy, in Knowledge-Based and Intelligent Information & Engineering Systems, 1999

[8] Tom White, Hadoop: The Definitive Guide, O'Reilly, 2009

[9] Zakir H. Ahmed, Genetic Algorithm for the Traveling Salesman Problem using Sequential Constructive Crossover Operator, in International Journal of Biometrics & Bioinformatics Volume (3): Issue (6)

[10] J. Dean and S. Ghemawat. Mapreduce: Simplified data processing on large clusters. Commun. ACM, 51(1):107–113, 2008.

[11] http://hadoop.apache.org

[12] Chuck Lam, Hadoop in Action, Manning, 2011

[13] http://www.iwr.uni-heidelberg.de/groups/comopt/software/TSPLIB95/